\documentclass[
   reprint,
   superscriptaddress,
   amsmath,amssymb,
   aps,
   prx,
   longbibliography,floatfix
]{revtex4-2}

\usepackage{braket}
\usepackage{graphicx}
\usepackage{amsmath}
\usepackage{amssymb}
\usepackage{float}
\usepackage[dvipsnames]{xcolor}
\usepackage[colorlinks=true,urlcolor=blue,linkcolor=blue,citecolor=blue]{hyperref}
\usepackage[capitalize]{cleveref}
\usepackage[margin=0.75in]{geometry}
\usepackage{physics}
\usepackage{bm}

\DeclareFontFamily{U}{wncy}{}
\DeclareFontShape{U}{wncy}{m}{n}{<->wncyr10}{}
\DeclareSymbolFont{mcy}{U}{wncy}{m}{n}
\DeclareMathSymbol{\Sh}{\mathord}{mcy}{"58} 

\usepackage{mathtools}
\DeclarePairedDelimiter{\ceil}{\lceil}{\rceil}

\begin{document}

\title{Compact Pulse Schedules for High-Fidelity Single-Flux Quantum Qubit Control}

\author{Ross~Shillito}
\affiliation{1QB Information Technologies (1QBit), Vancouver, British Columbia, Canada}
\affiliation{Institut quantique and D\'epartement de physique, Universit\'e de Sherbrooke, Sherbrooke, Qu\'ebec, Canada}

\author{Florian~Hopfmueller}
\affiliation{1QB Information Technologies (1QBit), Vancouver, British Columbia, Canada}
\affiliation{Department of Physics \& Astronomy, University of Waterloo, Waterloo, Ontario, Canada}
\affiliation{Perimeter Institute for Theoretical Physics, Waterloo, Ontario, Canada}

\author{Bohdan~Kulchytskyy}
\affiliation{1QB Information Technologies (1QBit), Vancouver, British Columbia, Canada}

\author{Pooya~Ronagh}
\email[Corresponding author: ]{pooya.ronagh@1qbit.com}
\affiliation{1QB Information Technologies (1QBit), Vancouver, British Columbia, Canada}
\affiliation{Institute for Quantum Computing, University of Waterloo, Waterloo, Ontario, Canada}
\affiliation{Department of Physics \& Astronomy, University of Waterloo, Waterloo, Ontario, Canada}
\affiliation{Perimeter Institute for Theoretical Physics, Waterloo, Ontario, Canada}

\begin{abstract}
In the traditional approach to  controlling  superconducting qubits using microwave pulses, the field of pulse shaping has emerged  in order to assist in the removal of leakage and increase gate fidelity. However, the challenge of scaling microwave control electronics has created an opportunity to explore alternative methods such as single-flux quantum (SFQ) pulses.  For qubits controlled by SFQ pulses, high fidelity gates can be achieved by optimizing the binary control sequence.  We extend the notion of the derivative removal by adiabatic
gate (DRAG) framework to a transmon qubit controlled by SFQ drivers. The proposed implementation of SFQ pulse sequences can be stored in 22 bits or fewer, with gate fidelities exceeding 99.99\%. This modest memory requirement could help reduce the footprint of the SFQ coprocessors and power dissipation while preserving their inherent advantages of scalability and cost-effectiveness.
\end{abstract}
\date{\today}

\maketitle

\section{Introduction}
Superconducting qubits are a promising architecture for quantum computation, demonstrating high fidelities, long coherence times, and scalability to hundreds of qubits~\cite{Blais2021,Kjaergaard2020}. However, scaling to thousands of qubits, required for fault-tolerant computation, remains an experimental challenge, with heat dissipation issues and rising costs~\cite{Krinner2019}. Using single-flux quantum (SFQ) pulses to control a qubit's state has been proposed as an alternative to traditional microwave drives~\cite{McDermott2014,McDermott2018}. In this scheme, voltage pulses whose integrated area equates to a superconducting flux quantum $\Phi = h/2e$ are delivered to the qubit~\cite{Mancini1999,Jao1995}, controlled by a classical coprocessor integrated into the chip. Such an architecture is very robust and expends significantly less power, leading to fewer heating issues and allowing for much greater scalability~\cite{Likharev1991,Nakajima1991}, making it a strong candidate for noisy, intermediate-scale quantum era devices. Further, all of the pulses irradiating the qubit are precisely equivalent, allowing for reliable control. However, due to quasiparticle poisoning, this digital control of qubits typically lags behind their microwave counterpart in precision, with experimental single-qubit gate fidelities ranging from 95\%--98\%~\cite{Leonard2019,Howe2022,Liu2023}. This limited fidelity is expected to be addressed in future iterations of the architecture~\cite{McDermott2018}. 

Traditionally, control of a qubit is conducting using a sequence of evenly spaced SFQ pulses, referred to as a ``pulse train''~\cite{Leonard2019}. While this scheme is simple and easy to implement, the coherent error in the fidelity of operations is generally dominated by leakage~\cite{Libermann2016}. Consequently, there has been an increase in literature on optimal schemes to control these qubits, with genetic and trust-region algorithms proposed~\cite{Libermann2016,Vogt2022}. Gates implemented with such optimized pulse sequences are predicted to have fidelities as high as $99.99\%$, on par with microwave gate methods. However, these sophisticated pulse sequences face higher memory requirements than pulse trains, as the pulses must be encoded as a binary sequence on a bit-shift register~\cite{Mukhanov1993}, which functions as the coprocessor memory. Minimizing  memory requirements is of paramount importance for SFQ-based technologies, resulting in a reduction in the power dissipation and along with it the footprint of the coprocessor~\cite{McDermott2018}. In order to reduce hardware requirements, it is possible to tailor schemes to be ``hardware efficient'', requiring fewer bits to encode a sequence, with proposed methods resulting in similar fidelities, but with a reduction in the bit requirements from 250 to fewer than 55~\cite{Kangbo2019}.

On most larger-scale superconducting architectures, such as  on state-of-the-art surface codes~\cite{Krinner2022,Acharya2023}, pulses optimized via gradient descent~\cite{Khaneja2005} are generally not used due to their complexity and the challenges of closed-loop optimization. Instead, the derivative removal by adiabatic gate (DRAG) pulse shaping technique is used ubiquitously for single- and two-qubit gates, due to its simple implementation and performance~\cite{Motzoi2009}. In this work, we propose a digital analog to the DRAG framework, which implements arbitrary single-qubit rotations with fidelities exceeding 99.99\%. As the majority of the pulses are simply an SFQ train, this greatly reduces the memory requirements for encoding the pulse sequences, with at most 22 bits required for implementation. Consequently, we expect a significant reduction in the hardware overhead for implementing high fidelity SFQ pulse schemes on this architecture.

The paper is organized as follows. In \cref{sec:Model}, we present a model for the SFQ pulses using a transmon superconducting qubit. In \cref{sec:DRAG}, we provide details on the discrete equivalent of DRAG pulses, with analogies to the continuous model, and discuss the encoding of the pulses in \cref{sec:Encoding_Pulse}. In \cref{sec:Gate_Opt_Ramps}, we report the results and fidelities of this pulse scheme and demonstrate its robustness against parameter variations. \cref{sec:Conclusion} concludes the paper.

\section{A Model for SFQ Control} 
\label{sec:Model}

We consider the model for the transmon qubit~\cite{Koch2007}
\begin{equation}
	\widehat{H}_0 = 4E_C(\hat{n}-n_g)^2 - E_J\cos{\hat{\varphi}}\,,
 \label{eqn:Transmon_Ham}
\end{equation}
where $E_C$ and $E_J$ are the charging and Josephson energies, respectively, and $n_g$ is the offset gate charge. In the regime $E_J \gg E_C$, the lowest-lying energy levels become exponentially insensitive to $n_g$; we thus take $n_g=0$ for simplicity.  This Hamiltonian can be expanded in the Fock basis, yielding the approximate model
\begin{equation}
    \widehat H_0 \approx \omega_q \hat b^\dagger \hat b - \frac{\alpha}{2}\hat b^{\dagger 2}\hat b^2,
\end{equation}
with the qubit frequency $\omega_q \approx \sqrt{8E_C E_J}$ (throughout this paper, we adopt the convention $\hbar=1$), the anharmonicity $\alpha \approx -E_C$, and $\hat b^\dagger$ and $\hat b$ are the creation and annihilation operators of the transmon, respectively. Unless otherwise mentioned, we assume that $\omega_q/2\pi = 5$ GHz and $\alpha/2\pi = -250$ MHz, corresponding to the ratio $E_J/E_C \approx 69$.

\subsection{Qubit Control by Continuous Driving}
Transmon qubits are traditionally driven using a charge line. The charge operator in the Fock basis can be represented as
\begin{equation}
    \hat n \approx \frac{i}{2}\left(\frac{E_J}{2E_C}\right)^{\frac{1}{4}}(\hat b^\dagger - \hat b) = ir(\hat b^\dagger - \hat b),
\end{equation}
where $r$ is the magnitude of the zero-point fluctuations of the charge operator. Time-dependent control of the transmon qubit can thus be represented with the Hamiltonian
\begin{equation}
    \begin{aligned}
    \widehat H(t) &=\widehat H_0 +\widehat H_\mathrm{D}(t),
  \label{eqn:microwavedrivenHamiltonian}
  \end{aligned}
\end{equation}
where $\widehat H_\mathrm{D}(t) = i\Omega(t)(\hat b^\dagger - \hat b)$ is the drive Hamiltonian and $\Omega(t)$ is a time-dependent pulse. The driving field is taken to be classical, thus assuming a weak coupling of the transmon to its input port \cite{Blais2007}. In general, this pulse consists of a slowly varying pulse envelope, and a fast carrier frequency targeting the qubit transition, $\Omega(t) = \Omega_0(t)\cos(\omega_qt + \phi)$, where $\phi$  determines the axis of rotation around the Bloch sphere and $\int_0^{t} \Omega_0(t') dt'$ is the angle through which the qubit state is rotated.
Then, entering a rotating frame  $\widehat U = \exp(i\omega_q \hat b^\dagger \hat b t)$ and assuming $\phi=0$, we find the Hamiltonian
\begin{equation}
    \widehat H'(t) = -\frac{\alpha}{2}\hat b^{\dagger 2}\hat b^2 +\frac{i}{2}\Omega_0(t)\left[\hat b^\dagger (1+e^{2i\omega_q t}) - \hat b(1+e^{-2i\omega_q t})\right].
    \label{eqn:HamRotatingFrame}
\end{equation}
Then, the fast oscillatory terms $e^{2i\omega_q t}$ are dropped, yielding a rotating-wave approximation (RWA) Hamiltonian
\begin{equation}
    \widehat H_\textrm{RWA}'(t) = -\frac{\alpha}{2}\hat b^{\dagger 2}\hat b^2 +\frac{i}{2}\Omega_0(t)(\hat b^\dagger - \hat b).
    \label{eqn:RWAHamiltonian}
\end{equation}
If the ratio $\Omega_0/\alpha$ is sufficiently small, the leakage level $|2\rangle$ of the transmon will be only weakly populated during the operation of the gate. Population of this leakage level can be reduced by pulse shaping, which is discussed in \cref{sec:DRAG}.

\subsection{Qubit Control Using SFQ Pulses}

\begin{figure}[h!]
    \centering
    \includegraphics[width=0.5\textwidth]{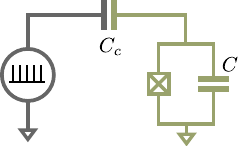}
    \caption{Transmon qubit (green) capacitively coupled to the SFQ generator. We approximate the arrival of the pulses at the qubit as Dirac delta functions, resulting in a pulse train.}
    \label{fig:Transmon_Qubit_SFQ}
\end{figure}

We now consider a different regime of qubit control: a sequence of discrete, fast pulses on the transmon qubit. \Cref{fig:Transmon_Qubit_SFQ} shows a circuit diagram of an SFQ-driven qubit. Whilst in practice the SFQ pulses arriving at the qubit have finite width ($\approx $1ps), this duration is negligible compared to the time between each SFQ pulse ($\approx $200ps), and as such, each pulse can be accurately modeled with a Dirac delta function~\cite{McDermott2014}. The corresponding driving Hamiltonian for such a pulse train can be taking the form
\begin{equation}
\hat H_{\textrm{sfq}} = -\frac{C_c}{C'}V(t)\hat n,
\end{equation}
where $V(t) = \Phi_0 \sum_n \delta(t - nT)$ is the voltage source. By recasting the charge operator's zero-point fluctuations in terms of the qubit frequency \cite{Koch2007} and computing the rotation corresponding to a single Dirac pulse, we arrive at the unitary
\begin{equation}
	\widehat{U}_\textrm{kick}(\theta) = \exp(-\frac{\theta}{2}\left[\hat b^\dagger - \hat b\right]),
\end{equation}
where the kick angle $\theta = C_c\Phi_0\sqrt{2\omega_q/C}$, in radians. Here, $C$ is the self-capacitance of the qubit, $C_c$ is the capacitance between the qubit and the SFQ driver, and $\omega_q$ is the qubit frequency~\cite{McDermott2014}. In this work, we consider a kick angle $\theta=0.03$, which is consistent with previous works and shown to be approximately optimal for counterbalancing coherent and non-coherent errors \cite{Libermann2016, Kangbo2019}. We fully anticipate this method holding for a larger range of kick angles where the RWA remains valid. 

The unitary evolution of the qubit can be described as a series of operations---the free evolution of the qubit and the SFQ kick operators:
\begin{equation}
	\widehat U = \prod_{i=0}^N \exp(-i\widehat H_0 t_i)\widehat U_{\textrm{kick}},
 \label{eqn:U_SFQ_Simple}
\end{equation}
where $N$ is the total number of kicks and $t_i$ is the time interval between the $i$-th and \mbox{$(i+1)$-th} kick. Within a two-level approximation, we note that 
\begin{equation}
	\exp(-i\widehat H_0 T)\widehat U_{\textrm{kick}} \approx \widehat R_Y(\theta).
\end{equation}
where the qubit period $T = 2\pi/\omega_q$. Consequently, by choosing all waiting times $t_i$ to be equal to the qubit period, we can approximate arbitrary $Y$ rotations around the Bloch sphere by applying a pulse train, where $\widehat R_Y(\theta_{\textrm{target}}) \approx \widehat R_Y(n \theta)$.

\subsection{Continuous Drive Equivalence to SFQ Kicks}

We show here that the SFQ and continuous pictures are equivalent within the RWA. We first recall the Dirac comb function
\begin{equation}
    \Sh_{\phi,T}(t) = \sum_{n=-\infty}^\infty \delta(t-nT - \phi),
\end{equation}
which approximates the SFQ pulse train, where $\phi$ is a phase offset. This results in the effective Hamiltonian
\begin{equation}
    \widehat H(t) = \widehat H_0 + \frac{i\theta}{2}\Sh_{0,(2\pi/\omega_q)}(t) (\hat b^\dagger - \hat b),
    \label{eqn:Relate_microwave_to_SFQ_amp}
\end{equation}
which, when exponentiated, yields~\cref{eqn:U_SFQ_Simple}, with all waiting times $t_i = 2\pi/\omega_q$. Making use of the identity
\begin{equation}
    \Sh_{\phi,(2\pi/\omega_q)}(t) \equiv \frac{\omega_q}{2\pi}\sum_{n=-\infty}^\infty e^{-in(\omega_q t+\phi)},
\end{equation}
and once again entering a rotating frame, yields the Hamiltonian
\begin{equation}
    \begin{aligned}
        \widehat H'(t) &= -\frac{\alpha}{2}\hat b^{\dagger 2}\hat b^2  + H_\mathrm{D}'(t),\\
        H_\mathrm{D}'(t) &= \frac{i\omega_q\theta}{4\pi}\hat b^\dagger \sum_{n=-\infty}^\infty e^{-in(\omega_q t+\phi)} + \textrm{H.c.}
    \label{eqn:Complex_U_Hamiltonian}
    \end{aligned}
\end{equation}
Once again making an RWA, this time ignoring terms oscillating at $\omega_q $ or faster, we return to the RWA form of the Hamiltonian in \cref{eqn:RWAHamiltonian}, and thus directly relate the drive amplitude and kick angle
\begin{equation}
    \Omega_0(t) = \omega_q\theta/2\pi,
    \label{eqn:Equiv_Drive_Amp}
\end{equation}
demonstrating the equivalence between the continuous drive and SFQ pictures.

\section{Leakage Removal via Ramps}
\label{sec:DRAG}

\begin{figure}[t!]
    \centering\includegraphics[width=0.5\textwidth]{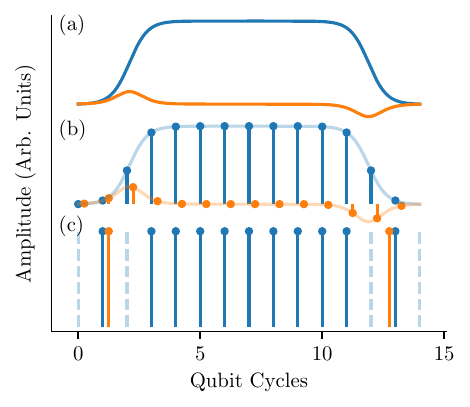}
	\caption{Example pulse shapes for implementing an $\widehat R_Y(\theta)$ gate in microwave and SFQ architectures. (a) Example $\textrm{tanh}$ pulse envelope (blue) microwave drive, with accompanying pulse derivative (orange), implemented on the opposing quadrature for DRAG. (b) Dirac delta approximation to the envelopes in (a), valid within the RWA. (c) Example SFQ pulse sequence, inspired by the sequence in (b). Here, the kick angle is fixed. Negative weighting on kicks can be effectively implemented by shifting the delta function by $T/2$.}
\label{fig:Pulse_Shape_SFQ}
\end{figure}

The DRAG framework is used ubiquitously in superconducting qubit architectures to mitigate the effects of leakage errors~\cite{Motzoi2009}. 
By augmenting a given pulse shape $\Omega(t)$ with its time derivative, leakage effects can be greatly reduced. Consider the pulse shape
\begin{equation}
    \label{eq:DRAG}
	\Omega'(t) = \Omega(t) +\frac{i c\dot\Omega}{\alpha} ,
\end{equation}
where $c$ is a scalar. \Cref{fig:Pulse_Shape_SFQ}(a) shows an illustration of a DRAG pulse, with the blue curve corresponding to the primary pulse $\Omega(t)$ and the orange curve to the derivative $\dot \Omega(t)$. The choice of $c=1$ effectively eliminates the spectral component of the pulse at the leakage frequency, as can be seen from the Fourier transform of the pulse over the gate time $T_\mathrm{g}$~\cite{Motzoi2013}:
\begin{equation}
    \begin{aligned}
        S(\Omega,\alpha) &=   \int_0^{T_\mathrm{g}} \Omega(t) e^{-i\alpha t} dt = -i\int_0^{T_\mathrm{g}} \frac{\dot\Omega(t)}{\alpha} e^{-i\alpha t} dt.
    \end{aligned}
    \label{eqn:SpectralComponent}
\end{equation}
The relative phase difference $(-i)$ between the two terms in \cref{eqn:SpectralComponent} is equivalent to a $\pi/2$ phase shift in the driving field:
\begin{equation}
	\widehat H(t) = \widehat H_0 + i(\hat b^\dagger - \hat b) \left(\Omega(t)\cos(\omega_q t) + \frac{c\dot\Omega(t)}{\alpha}\sin(\omega_q t) \right).
    \label{eqn:Microwave_Drive_Ham_DRAG}
\end{equation}

We consider a mapping from the DRAG formalism to an equivalent SFQ interpretation. As is standard practice, we assume the drive pulse $\Omega(0)=\Omega(T_\mathrm{g}) = 0$ is zero at the start and end of the gate, with gate time $T_\mathrm{g}$. Within the RWA made in the previous section, the drive Hamiltonian is equivalent to 
\begin{equation}
    \begin{aligned}
        &\widehat H_\mathrm{D}(t) =\frac{i}{2}(\hat b^\dagger - \hat b)\\
        &\times\left[ \Omega_y(t)\Sh_{0,2\pi/\omega_q}(t) +\Omega_x(t) \Sh_{\pi/2, 2\pi/\omega_q}(t)\right],\\
    \end{aligned}
    \label{eqn:Equiv_RWA_H}
\end{equation}
where
\begin{equation}
    \Omega_y(t) = \frac{2\pi\Omega(t)}{\omega_q}\textrm{\,\, and \,\,}\Omega_x(t) = \frac{2c\pi\dot\Omega(t)}{\alpha\omega_q}.
\end{equation}
We now consider the unitary corresponding to what is known as an ``on-ramp'', where in the case of the microwave drive the amplitude is slowly increased. The equivalent SFQ evolution operator thus becomes
\begin{equation}
    \begin{aligned}
        \widehat U_\textrm{on}(t_\mathrm{r}) &= \mathcal{T}\exp(-i\int_0^{t_\mathrm{r}} [ \widehat H_0 + \widehat H_\mathrm{D}(t') ] dt')\\
        &= \prod_n e^{-i3T/4}\widehat U_{\textrm{kick}}[\Omega_x(nT + T/4)]\\
        &\ \ \ \ \times e^{-iT/4}\widehat U_{\textrm{kick}}[\Omega_y(nT)],
    \end{aligned}
    \label{eqn:Equiv_RWA_SFQ}
\end{equation}
where $t_\mathrm{r}$ refers to the ``ramp time'', or the time it takes the pulse amplitude to reach its maximum value, $t_{\mathrm{r}} = \arg \max_t \Omega(t)$. An example pulse sequence corresponding to this unitary is given in \mbox{\cref{fig:Pulse_Shape_SFQ}(b),} where each of the SFQ kicks takes on a different effective amplitude.

While the SFQ Hamiltonian in \cref{eqn:Equiv_RWA_H} is equivalent to the continuous drive in \cref{eqn:Microwave_Drive_Ham_DRAG} up to an RWA, we require all the SFQ kick angles $\theta$ to be identical, as this is fixed by the circuit capacitances~\cite{Leonard2019}. As such, we approximate the DRAG condition introduced in \cref{eq:DRAG} by first equating the integrated pulses:
\begin{equation}
    \begin{aligned}
        \Omega_x &= c\frac{\dot \Omega}{\alpha} \rightarrow \int_0^{t_\mathrm{r}} \Omega_x(t) dt = \frac{c}{\alpha}{\Omega(t_\mathrm{r})}.
    \end{aligned}
\end{equation}
We quantize this relation, with $\int_0^{t_\mathrm{r}} \Omega_x(t) dt = n\theta$, where $n$ is the effective number of pulses producing qubit rotation about the $x$ axis. Using the relation in \cref{eqn:Equiv_Drive_Amp}, we describe the target number of kicks timed to produce the $x$-axis rotation as 
\begin{equation}
    n = \frac{c}{\alpha}T.
    \label{eqn:SFQ_DRAG}
\end{equation}
The goal is thus to find ``SFQ ramps'', or specific sets of SFQ pulses, that implement \cref{eqn:SFQ_DRAG}.  In this sense, the evolution of the system will be decomposed into three pieces: the on-ramp evolution, an SFQ ``pulse train'' corresponding to a set of SFQ kicks, and an ``off-ramp'' evolution, as depicted in \mbox{\cref{fig:Pulse_Shape_SFQ}(c):}
\begin{equation}
    \widehat U(T) = \widehat U_\textrm{off}(t_\mathrm{r}) \widehat U_\textrm{train}\widehat U_\textrm{on}(t_\mathrm{r}).
    \label{eqn:U_SFQ_evolution_total}
\end{equation}
In close analogy with analog DRAG pulses, the off-ramp $\widehat U_\textrm{off}(t_\mathrm{r})$ should be symmetric with respect to the on-ramp unitary along the $y$-axis and antisymmetric along the \mbox{$x$-axis} for the implementation of an $\widehat R_Y(\theta)$ gate. While we cannot control the sign of the amplitude of the kick, the equivalent $\widehat R_X(-\theta)$ rotation can be achieved by an effective shift of the arrival time of the pulses in the $x$ direction by $T/2$. 

While the choice of $c=1$ minimizes leakage, it has been demonstrated that $c=0.5$ maximizes the gate fidelity due a reduction in phase errors~\cite{Lucero2010}. Alternatively, DRAG can be coupled with a virtual $\widehat Z$ gate to suppress both phase and leakage~\cite{McKay2017,Christopher2018}. As the choice of the $c$ parameter is limited by the quantization relation in \cref{eqn:SFQ_DRAG} and we cannot correct for arbitrary phase errors with a virtual $\widehat Z$ gate, we simply optimize the gate fidelity, as discussed in \cref{sec:Gate_Opt_Ramps}.

\section{Encoding the Pulse Sequence}\label{sec:Encoding_Pulse}

The possible choices of SFQ ramps will depend on the SFQ global clock, which controls the arrival time of the pulses. Thus, to achieve the resonant pulse train, the SFQ clock frequency must be some integer multiple of the qubit frequency. We begin by assuming a clock frequency of 20 GHz, such that the period of the clock $T_\mathrm{c}= T/4$, which, as discussed in the previous section, is required for implementing the equivalent evolution to continuous dynamics. As such, by choosing pulses every four clock cycles, we can effectuate $\widehat R_Y(\theta)$ rotations. Let us adopt the binary notation $ i j k l$, where $\lbrace i,j,k,l \rbrace \in \lbrace 0,1\rbrace$, to represent a sequence that is sent to the SFQ driver. Each $1$ and $0$ refers to either the presence or absence of an SFQ kick, followed by a waiting time of $T/4$. 
As a simple example, the binary sequence $1 0 0 0$ effectuates
\begin{equation}
    1 0 0 0 \rightarrow e^{-i\widehat H_0 T} \widehat U_{\textrm{kick}} = \widehat R_Y(\theta)
\end{equation}
on the qubit subspace. Similarly, the choices $0 1 0 0$, $ 0 0 1 0$, and $0 0 0 1$ effectuate $\widehat R_X(-\theta)$, $\widehat R_Y(-\theta)$, and $\widehat R_X(\theta)$ rotations, respectively. To demonstrate this more concretely, we consider the simplest possible SFQ ramp of $0 1 0 0$. The evolution corresponding to \cref{eqn:U_SFQ_evolution_total} can thus be represented by the binary sequence (recalling that the binary sequence is read from left to right)
\begin{equation}
    \begin{aligned}
    &\widehat U(T) \longrightarrow (0 1 0 0) + (1000)^{(N-1)} + (1) + (0  0 1 0)\\
    &= \widehat R_Y(-\theta) \widehat R_Z(\pi/2)\widehat R_Y(\theta) \widehat R_Y(\left[N-1\right]\theta) \widehat R_X(-\theta) \\
    &= \widehat R_X(\theta)\widehat R_Y(N\theta) \widehat R_X(-\theta),
    \end{aligned}
\end{equation}
up to an $\widehat R_Z (\pi/2)$ rotation. 
To consider more-complex ramps, we concatenate up to a total of five binary sequences from the subset
$(0000,\ 1000,\ 0100,\ 1100)$, which correspond to $\widehat I, \widehat R_Y(\theta), \widehat R_X(-\theta)$, and $\widehat R_X(-\theta)\widehat R_Y(\theta)$, respectively. Consequently, a choice of $n$ qubit cycles for the ramp leads to a total of $4^n$ possible ramps. The ignored binary sequences implement rotations in the opposite direction indicated by the DRAG technique or against the targeted gate rotation. In \cref{app:Fourier_Component}, we demonstrate that the optimized DRAG sequences greatly reduce the spectrum of the pulse sequence at the leakage frequency, in agreement with \cref{eqn:SpectralComponent}.

An advantage of SFQ hardware is the speed and accuracy at which pulses can be delivered to the qubit. Consequently, global clocks of 40 GHz or higher are feasible~\cite{Yamanashi_2021,Chen1998}. For the provided choice of parameters, such a clock frequency would equate to eight times the qubit frequency, allowing for greater control of the qubit. In line with the above terminology, we additionally consider ramps for the ``8$\times$'' clock. There are eight cycles to consider: 
\begin{align*}
        &0 0 0 0 0 0 0 0,\  1 0 0 0 0 0 0 0,\  0 1 0 0 0 0 0 0,\ 0 0 1 0 0 0 0 0,\\
        &1 1 0 0 0 0 0 0,\  1 0 1 0 0 0 0 0,\  0 1 1 0 0 0 0 0,\ 1 1 1 0 0 0 0 0.
\end{align*}
The faster clock enables pulses, such as $0 1 0 0 0 0 0 0$, that result in an effective $\theta/\sqrt{2}$ kick in both the $x$ and $y$ quadratures, giving faster and more-precise control.

In the case of the proposed ramps, a significantly smaller number of bits is required than traditional methods~\cite{Libermann2016, Vogt2022}. Given there are four choices of unitary rotation applied at each qubit cycle for the 4$\times$ clock and eight for the 8$\times$ clock, the full on-ramp sequence can be described by $n \times \log_2(4)$ and $n\times \log_2(8)$ bits, respectively, where $n$ is the number of qubit cycles in the on-ramp. The number of pulses in the pulse train $N$ could easily be represented in binary form with $\ceil{\log_2 N}$ bits, and the encoding for the off-ramp can be recycled from the on-ramp. We also assume that no additional programmable memory is required to store bit sequences used in the train portion of the pulse as these can be hard-coded and shared among all the qubits. Consequently, a five-ramp cycle with 84 pulses in the pulse train can be implemented using only $5 \times \log_2(4) + \ceil{\log_2 84}=17$ bits for the 20 GHz clock. With the 40 GHz clock, the same gate requires  $5 \times \log_2(8)+ \ceil{\log_2 84}=23$ as an additional bit is required to encode each rotation in the ramp sequence while the number of pulses in the ramp and train remain the same.  As shown in \cref{sec:Gate_Opt_Ramps}, this is sufficient for implementing gates with greater than $99.99\%$ fidelity. By sharing the encoding of the ramps and pulse trains between neighbouring qubits in a larger-scale device, the effective bit requirement per qubit could be brought even lower using this technique.
\section{Gate Optimization Using Ramps}
\label{sec:Gate_Opt_Ramps}

We now demonstrate the performance of the proposed SFQ pulse trains with appropriate ramps for relevant target angles. We restrict our analysis to targeting arbitrary $y$ rotations, with target operators of the form
\begin{equation}
	\widehat U_{\textrm{target}} = \cos(\theta_\textrm{target}/2) \widehat I - i\sin(\theta_\textrm{target}/2) \widehat Y.
\end{equation}
We note that the equivalent $\widehat R_X(\theta)$ rotations can be achieved using $(0001)^N$ pulse trains.  To quantify the gate performance in the presence of leakage, we use the average gate fidelity metric in the absence of a loss channel~\cite{Christopher2018}:
\begin{equation}
    \overline F(\mathcal{E}) = \frac{2F_{\textrm{pro}}(\mathcal{E}) + 1 - L_1}{3}.
    \label{eqn:AvgGateFidelity}
\end{equation}
Here, the process fidelity is defined as
\begin{equation}
    F_\textrm{pro}(\mathcal{E}) = \frac{1}{4}\left|\Tr[\widehat U^\dagger _{\textrm{target}} \widehat U_Q]\right|^2,
\end{equation}
and the leakage metric is
\begin{equation}
    L = 1-\frac{1}{2}\Tr(\widehat U_Q \widehat U^\dagger_Q),
\end{equation}
where $\widehat U_Q$ is the projection onto the computation subspace of the full unitary operation $\widehat U$ (see \cref{eqn:U_SFQ_evolution_total}), which takes into account the higher-energy levels and leakage effects. To ensure the numerical accuracy of our work, we consider the full-cosine Hamiltonian of the transmon in \cref{eqn:Transmon_Ham}, diagonalized in the charge basis with 200 states and projected onto the lowest seven eigenstates. This ensures that the simulations will capture additional leakage to the $|3\rangle$ state and higher transmon eigenstates. Further, the kick operator is taken as the exponential of the charge operator:
\begin{equation}
    \widehat{U}_\textrm{kick}(\theta) = \exp(-\frac{\theta'}{2} \hat n),
\end{equation}
where $\theta' = \theta/r$, accounting for the zero-point fluctuations of the charge operator $\hat n$. We consider the performance of this method with a finite qubit lifetime in \cref{app:ME_Sims} with \cref{fig:Ramp_Pulse_Profiling} detailing the results of this analysis.

\begin{figure}[h]
    \centering
    \includegraphics[width=0.5\textwidth]{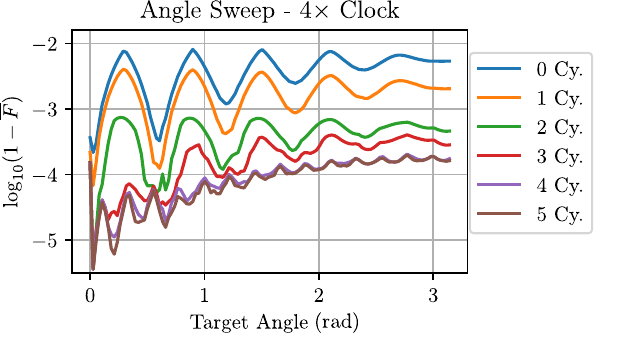}
    \caption{Maximum fidelity of target rotation operations $\widehat R_Y(\theta)$ as a function of the target angle (in radians). The optimal pulse train length and optimal ramp for each ramp length is postselected.} 
    \label{fig:Angle_Sweep}
\end{figure}

To find the optimal ramps, we perform an exhaustive search over relevant ramps that maximize $\overline F(\mathcal E)$ for each target gate $\widehat R_Y(\theta)$. As we consider a maximum of five cycles, all possible ramps can easily be tested. For longer ramps and more-complex systems, a combination of the constraint in \cref{eqn:SFQ_DRAG} and the pulse sequence spectrum (see \cref{app:Fourier_Component}, in particular, \cref{fig:FourierTransform})) can be used to reduce the possible ramp subspace and accelerate the optimization.  In \cref{fig:Angle_Sweep}, we plot the fidelity of operation of a target rotation $\widehat R_Y(\theta)$ as a function of the target angle for the 4$\times$ clock. The different colours represent differing numbers of on-ramp and off-ramp cycles, with blue (0 cycles) representing only a pulse train. As expected, the longer ramps yield significantly higher fidelities, with four cycles yielding less than an error of $10^{-4}$ for most target angles. However, we note that the fidelity is eventually limited for four cycles, with almost identical fidelities for five cycles.

\begin{figure*}[t!]
    \centering
    \includegraphics[width=0.8\textwidth]{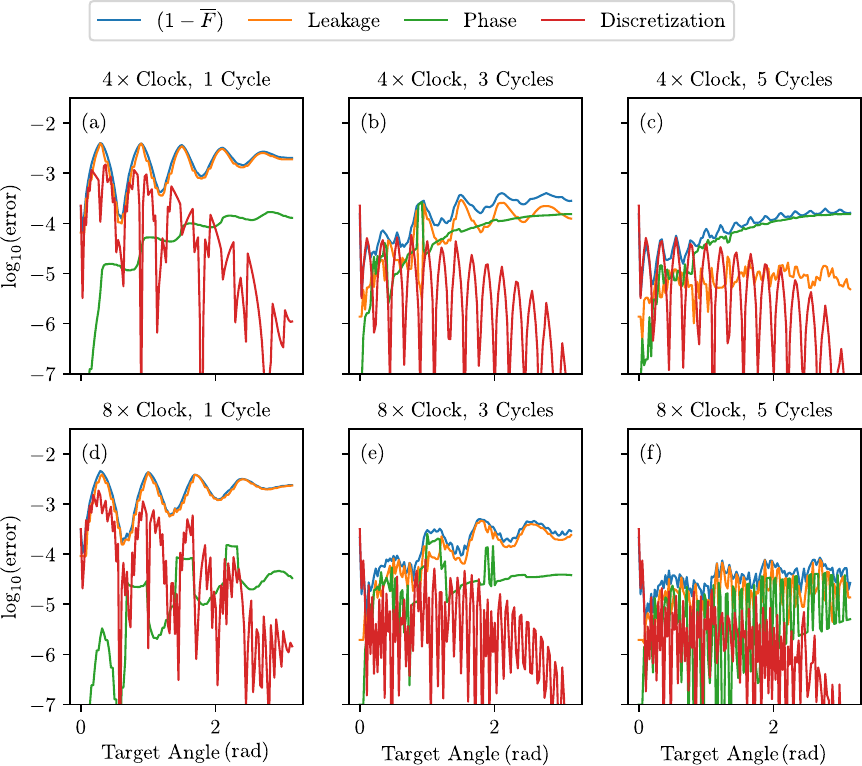}
    \caption{Error analysis for the 4$\times$ clock (a)--(c) and 8$\times$ clock (d)--(f). For the shorter ramps, the leakage remains the dominant source of error, but is significantly suppressed with a five-cycle on-ramp. For small target angles, the discretization error dominates. For larger angles, the phase error dominates.}
    \label{fig:Big_Error_Analysis}
\end{figure*}

In order to gain greater insight into the source of the errors, it is useful to perform a decomposition of the projected operator $\widehat U_Q$ into the Pauli matrices:
\begin{equation}
    \begin{aligned}
	\widehat U_Q &= 
  (1-\delta)(c_I \widehat I + c_X \widehat X + c_Y \widehat Y + c_Z \widehat Z),
    \end{aligned}
\end{equation}
where $\sum_i |c_i|^2 = 1$ and $ L_1 = 2\delta -\delta^2$, with $\delta$ being an alternative way of characterizing the leakage. Here, we define two types of errors:
\begin{itemize}
    \item Discretization error---the error arising from having a kick angle incommensurate with the target angle. We define this error as
    \begin{equation}
        \mathcal{E}_{\textrm{discrete}}  =  \left|c_Y - \sin(\theta_\textrm{target}/2)\right|^2.
    \end{equation}
    \item Phase error---the error caused by virtual transitions to leakage levels. We define this error as
    \begin{equation}
        \mathcal{E}_{\textrm{phase}}  =  \left|{c_Z}\right|^2.
    \end{equation}
\end{itemize}
The contribution from $|c_X|^2$ is negligible due to the symmetry of the pulse.

We plot the types of errors for gates of an arbitrary angle $\theta_\textrm{target}$ for one cycle in \cref{fig:Big_Error_Analysis}(a) and (d), three cycles in \cref{fig:Big_Error_Analysis}(b) and (e), and five cycles in \cref{fig:Big_Error_Analysis}(c) and (f). Most pertinently, we see how the types of errors change as a function of the number of ramps and the target angle. First, we note the large oscillations in errors for the one-cycle case, which is accounted for almost entirely by coherent population and depopulation of the leakage level $|2\rangle$. These oscillations depend on both the chosen kick angle $\theta$ and the qubit's anharmonicity $\alpha$. As we include additonal ramps, the leakage is greatly suppressed (see \cref{app:Fourier_Component} and \cref{fig:FourierTransform} for further details). We finally note that, for the longest ramps, it is the phase-shift error that dominates, which accumulates through virtual transitions to the leakage level.

The introduction of the faster 8$\times$ clock allows for significantly greater control that can be exploited for further reduction in errors. For example, the 8$\times$ clock allows for a significant reduction in the phase shift error, as shown in \cref{fig:Big_Error_Analysis}(f), where the five-cycle ramp results in an improvement for up to an order of magnitude in the gate fidelity for a large fraction of target angles. As the average gate fidelity is targeted, the optimization alternates between leakage and phase errors, depending on which dominates. In analogy with Ref.~\cite{Christopher2018}, this would allow for a choice of the DRAG ramp to target the reduction in either leakage, phase, or average fidelity errors.

\section{Conclusion and Future Work}
\label{sec:Conclusion}

We have demonstrated a simple implementation of a discretized DRAG method for reducing gate errors caused by transitions into non-computational levels  in a transmon qubit. In the case of the SFQ pulses being delivered to the qubit at four times the qubit's frequency, the method is primarily limited by the phase-shift error induced by the leakage levels, but this limitation is mitigated with a faster external clock. Unlike possible alternative approaches for addressing phase errors, such as via SFQ clock frequency detuning \cite{Motzoi2013} or flux-tunable transmons, our method imposes minimal requirements on control electronics and qubit architecture. Moreover, due to the construction of the pulse, its hardware encoding is greatly compressible. For example, a 22-bit pulse allows us to obtain high-quality single-qubit gates reaching a fidelity greater than 99.99\%. Given the strong relationship between continuous DRAG pulses and SFQ sequences, we fully anticipate that more-sophisticated continuous methods seen in the literature, for example in Refs. \cite{Gambetta2011, Vezvaee2023}, could be translated to further improve the SFQ schemes in a hardware efficient manner. We further expect utilizing higher effective clock speeds could result in the design of more sophisticated ramps. Furthermore, hardware storage overheads could be potentially reduced by ramp sharing between different gates and qubits. The simplicity of the pulse encoding could make our scheme compatible with a closed-loop optimization where a subset of optimized pulses would be further calibrated directly on the hardware. We additionally anticipate that the method will generalize to other qubit architectures and gates, for example, two-qubit gates~\cite{torosov2024optimization,Jokar2021} controlled with flux pulses~\cite{Caldwell2018} or with couplers~\cite{Moskalenko2022}. The simplicity of our scheme allows us to move a step closer towards a vision of a scalable quantum chip controlled using SFQ technology.

\section*{Acknowledgements}

We thank our editor, Marko Bucyk, for his careful review and editing of the manuscript. We also thank Boyan Torosov for thoroughly proofreading the manuscript. We acknowledge Caleb Jordan, Oleg Mukhanov and Matthew Hutchings for useful discussions. During this work, R.~S. was a student at the Universit\'e de Sherbrooke and received funding through Mitacs. P.~R. acknowledges the financial support of Mike and Ophelia Lazaridis,
Innovation, Science and Economic Development Canada (ISED), and the Perimeter
Institute for Theoretical Physics. Research at the Perimeter Institute is supported in part by the Government of
Canada through ISED and by the Province of Ontario through the Ministry of
Colleges and Universities.

\newpage
\addcontentsline{toc}{section}{References}
\bibliography{refs}{}

\clearpage
\appendix

\section{Master Equation Simulations}
\label{app:ME_Sims}

\begin{figure*}
    \centering
    \includegraphics[width=\textwidth]{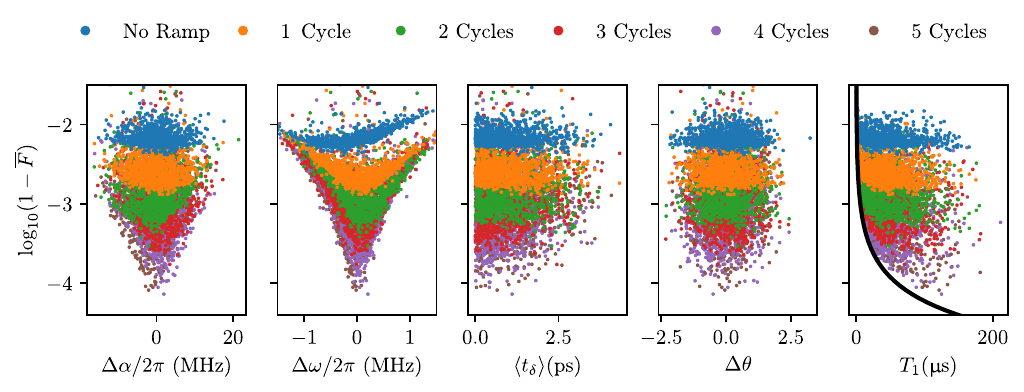}
    \caption{Performance of the SFQ pulse sequence as a function of the parameter variations for the 8$\times$ clock. We allow for deviations in the qubit anharmonicity $\alpha$, the frequency $\omega$, the jitter $\langle t_\delta \rangle$, and the kick angle $\theta$. We also include the effects of varying $T_1$, or, equivalently, the qubit decay rate $\gamma$.}
    \label{fig:Ramp_Pulse_Profiling}
\end{figure*}

To quantify the gate performance in the presence of leakage, timing jitter, and other possible errors, we use a more-general average gate fidelity metric~\cite{Christopher2018}:
\begin{equation}
    \overline F(\mathcal{E}) = \frac{2F_{\textrm{pro}}(\mathcal{E}) + 1 - L_1}{3},
    \label{eqn:AvgGateFidelityAdv}
\end{equation}
where the process fidelity is defined as
\begin{equation}
    F_\textrm{pro}(\mathcal{E}) = \frac{1}{4}\Tr[S^\dagger _G (P_2\otimes P_2)(S_\mathcal{E})],
\end{equation}
where $P_2S_\mathcal{E} (P_2)$ is the superoperator representation of the computed map projected onto the computational subspace, and $L_1 = 1-\Tr[P_2 S_\mathcal{E}(P_2)]/2$ quantifies the leakage of the channel.

For our simulations, we use a standard Lindbladian master equation:
\begin{equation}
    \dot \rho = -i[\widehat H(t),\rho] + \gamma \mathcal{D}[\widehat n_t^\Delta]\rho\,,
\end{equation}
where $\widehat n_t^\Delta$ is the upper-triangular matrix of the charge operator $\widehat n_t$. Translating this into the notation of SFQ kicks yields an effective evolution of the density matrix
\begin{equation}
     \rho(t) = \left[\prod_i (\widehat U_{\textrm{kick}} \cdot \widehat U^\dagger_{\textrm{kick}}) \exp(\mathcal{L}_0 (t_i+ t_{i,\delta})) \right]\rho(0),
\end{equation}
where $\mathcal{L}_0 = -i(\widehat H\cdot - \cdot \widehat H) + \gamma \mathcal{D}[\widehat n_t^\Delta] \cdot$. Here, we include an additional error, the jitter time $t_{\delta,i}$. For each simulation, we suppose that the jitter time is sampled from a normal distribution with a standard deviation $\sigma = \langle t_\delta \rangle$, which is varied for each simulation. Further, we suppose that the kick angle $\theta$ is allowed to deviate by $\Delta \theta/1000$. The deviations in frequency and anharmonicity are centred around $\omega/2\pi = 5$ GHz, $\alpha/2\pi = 250$ MHz.  

To test the robustness of the pulse train, we perform 1000 example simulations for each ramp length, with varying qubit parameters for the same chosen set of SFQ pulses. The results for the 8$\times$ clock are shown in \cref{fig:Ramp_Pulse_Profiling}. 

Here, we restrict our focus to the $\widehat R_Y(\pi)$ gate, but have found similar results for other target angles. The samples are normally distributed around the parameters with which the  gate is optimized.

We note that for the ``no-ramp'' evolution, the optimal frequency of the qubit is off-centre from the integer multiple of the clock. This is due to the phase shift induced by the virtual transitions to the leakage levels. As the number of ramps increases, the deviations become increasingly centred on the qubit frequency $(\Delta\omega = 0)$, and the fidelity continues to improve.

\section{Spectral Components of the Pulse Sequences}
\label{app:Fourier_Component}

\begin{figure}[ht]
    \centering
    \includegraphics[width=0.5\textwidth]{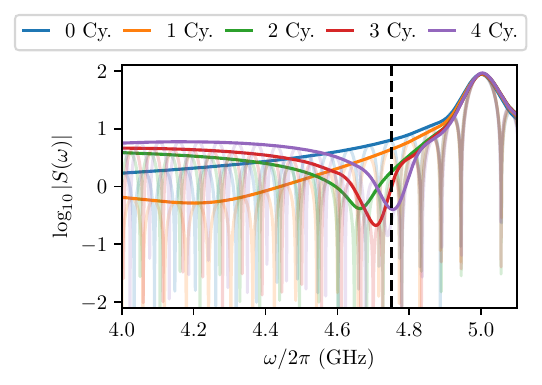}
    \caption{Fourier transform of a sequence of 87 pulses, beginning with no ramp (0 qubit cycles, shown in blue) and continuing up to a ramp of four qubit cycles (purple), optimized to implement an $\widehat R_Y(\pi)$ gate for the 20 GHz clock. The leakage frequency is indicated by a dashed line. The transparent lines indicate the true Fourier transform, while the opaque lines interpolate between each peak for clarity. As the number of ramps increases, the dip in spectral intensity is shifted with increasing accuracy to the leakage transition frequency.}
    \label{fig:FourierTransform}
\end{figure}

To confirm that a given pulse sequence matches the behaviour of an equivalent DRAG pulse, we can analyze the spectrum of the pulse being delivered to the qubit. We begin by representing the pulse sequence for a \mbox{20 GHz} external clock as a sum of Dirac delta functions,
\begin{equation}
    f(t) = \sum_m \delta(t - mT/4),
\end{equation}
where the indices $m$ are set by the number of pulses in the pulse train and the chosen ramp. We then take the Fourier transform, yielding
\begin{equation}
    S(\omega) = \sum_m e^{2i\pi m\omega/4}.
    \label{eqn:Spectral_Intensity}
\end{equation}
In \cref{fig:FourierTransform}, we plot the spectrum \cref{eqn:Spectral_Intensity} as a function of frequency for different ramp lengths, where the ramps are chosen so as to optimize the gate fidelity as described in \cref{sec:Gate_Opt_Ramps}. As can be seen, the increasing length of the ramp steadily decreases the spectral intensity at the leakage frequency (indicated by the dashed line) as the ramp length increases. For four cycles, there is a reduction of more than an order of magnitude in the spectral component at this frequency, confirming that the SFQ ramps indeed reduce the spectral contribution at the leakage transition frequency. This method could be used to analyze possible ramps independently of unitary simulations, and thus identify candidates for high-fidelity ramps without the need to explore the entire search space.

\end{document}